# TECHNIQUE FOR THE DRY TRANSFER OF EPITAXIAL GRAPHENE ONTO ARBITRARY SUBSTRATES

Joshua D. Caldwell,[1] Travis J. Anderson,[1] James C. Culbertson,[1] Glenn G. Jernigan,[1] Karl D. Hobart,[1] Fritz J. Kub,[1] Marko J. Tadjer,[2] Joseph L. Tedesco,[1] Jennifer K. Hite,[1] Michael A. Mastro,[1] Rachael L. Myers-Ward,[1] Charles R. Eddy Jr.,[1] Paul M. Campbell[1] and D. Kurt Gaskill[1]

[1]U.S. Naval Research Laboratory, 4555 Overlook Ave, S.W., Washington, D.C. 20375
[2]Electrical Engineering Department, University of Maryland, College Park, MD 20742

Joshua.caldwell@nrl.navy.mil

**RECEIVED DATE (to be automatically inserted after your manuscript is accepted if required according to the journal that you are submitting your paper to)**

Abstract: In order to make graphene technologically viable, the transfer of graphene films to substrates appropriate for specific applications is required. We demonstrate the dry transfer of epitaxial graphene (EG) from the C-face of 4H-SiC onto $SiO_2$, GaN and $Al_2O_3$ substrates using a thermal release tape. We further report on the impact of this process on the electrical properties of the EG films. This process enables EG films to be used in flexible electronic devices or as optically transparent contacts.

Reports of single layer graphene have drawn significant interest due to its exciting properties, such as ballistic carrier transport,[1,2] high thermal and electrical conductivity, optical transmission,[3-5] and high mechanical hardness.[6] Films used in those studies were created primarily via the exfoliation method,[1,5] which produces small-area flakes with variable size, shape and thickness. In order to use graphene as a conductive, optically transparent contact, the reproducible transfer of large-area graphene films that could subsequently be patterned into top-side contacts is required. Large-area graphene films can be produced by either metal-catalyzed growth on films of nickel[7-9] or copper[10] or by epitaxial growth through the sublimation of silicon from the surface of silicon carbide (SiC).[11-16] Epitaxial graphene (EG) films grown on the carbon-terminated surface of SiC have been shown to be of high quality, with room temperature Hall mobilities up to 4,200 $cm^2$/Vs having been observed in 16 mm x 16 mm C-face EG films grown within our lab.[17] Furthermore, we have observed room temperature Hall effect mobilities as high as 23,000 $cm^2$/Vs from specific 10 μm Hall crosses fabricated on C-face grown EG films. However, the C-face films, while exhibiting significantly higher mobilities than EG films grown on the Si-face, do tend to have a high variability in the EG film thickness across a given sample.

Here we present results illustrating a dry transfer technique using Nitto Denko Revalpha thermal release tape that has enabled the transfer of large areas (squares up to 16 mm on a side) of EG from the C-face SiC 'donor' substrate onto $SiO_2$ on Si 'handle' substrates. For this study C-face grown EG material was chosen due to its high carrier mobility. Successful transfers of C-face EG onto GaN and $Al_2O_3$ films will also be discussed, demonstrating that this process can be readily implemented for use with other handle substrates of interest. To date, the only reported transfer of EG from SiC utilized the exfoliation method,[18] producing sub-micron-sized samples, which are not amenable to large-scale device manufacturing. While a wet chemical approach has been successfully used to transfer metal-catalyzed graphene,[7-9] this process is not amenable to EG on SiC, as SiC is highly resistant to chemical etchants. We further report on Hall effect measurements that were performed on both the large-area transferred EG films and fabricated van der Pauw devices to determine the influence of the transfer process and subsequent lithography on the electrical properties of the transferred EG films. Raman spectroscopy was used to verify the quality and thickness of the EG films and X-ray photoemission spectroscopy (XPS) measurements were performed to verify the transfer efficiency. In addition, the XPS characterization of EG films during the transfer process revealed the presence of atomic silicon. These measurements all show that this transfer process can create large-area EG films on arbitrary substrates suitable for both device fabrication and further experiments exploring the impact of the substrate electrical properties upon the electrical behavior of the EG films.

EG was grown for these transfer experiments via the sublimation process on 2" C-face 4H-SiC substrates that were previously chemically-mechanically polished.[15, 16] The substrates were placed within an Aixtron/Epigress VP508 Hot-Wall chemical vapor deposition reactor and the surface was further prepared via an $H_2$ etch at 100 mbar and $1600^oC$ for 20 min. The chamber was evacuated to $(1.4-17)\times 10^{-4}$ mbar and the temperature was lowered to $1550^oC$ for EG formation. EG growth was carried out for 1 hour. After growth, the chamber was allowed to cool overnight.[16] Square samples 10 mm on a side were cut from the 2" C-face 4H-SiC substrate following EG growth, to enable multiple transfer attempts. The $SiO_2$ on Si 'handle' substrates used for most EG transfers consisted of 100 nm of thermally-grown $SiO_2$ on n-type Si.

In order to improve the strength of bonding between the transferred EG and the $SiO_2$ surface, a cleaning and surface preparation procedure was used to produce a hydrophilic surface on the $SiO_2$. The $SiO_2$ surface was cleaned using a 750 W $O_2$ plasma treatment for 5 min in a Plasma-Preen II-973, followed by ultrasonic SC1 cleaning (5:1:1, $H_2O:NH_4OH:H_2O_2$) at $40^oC$ for 14 min, ending with a 1 min rinse in megasonic water provided by a Honda Electronics PulseJet (W-357-3MP) system at full power. Immediately before the transfer, the handle substrates were treated with a 30 s, 750 W $O_2$ plasma treatment, a 1 min ultrasonic SC1 clean and a 1 min megasonic rinse to ensure a hydrophilic $SiO_2$ surface was produced. With the exception of a standard solvent clean, the EG surface did not require any specific preparation procedures for a successful transfer. For the EG transfers to p- and n-type GaN, 2 μm thick GaN films were grown on 2-inch, a-plane sapphire substrates using metal organic chemical vapor deposition (MOCVD). The deposition was initiated with a 25 nm AlN nucleation layer grown at $680^oC$ and 50 Torr with the subsequent GaN growth deposited at $1025^oC$ and 50 Torr. The n- and p-type doping was accomplished with disilane and $Cp_2Mg$, respectively. EG films were also transferred onto 100 nm $Al_2O_3$ films grown via atomic layer deposition (ALD) on a 20 mm x 20 mm double-side polished, epi-grade, c-face sapphire substrate at a chamber temperature of $300^oC$ using tetramethyl aluminum and water as the alternating precursors. For the EG transfer to both GaN and $Al_2O_3$, the surface preparation of the handle substrate consisted of only a 5 min, 750 W $O_2$ plasma treatment in the Plasma Preen system.

Following the prescribed substrate pretreatment procedures, a precut piece of Nitto Denko Revalpha Thermal Release Tape (Part No. 3193MS, 7.3 N/mm) was placed on the EG surface. The tape/EG/SiC sample 'stack' was then placed on a silicon wafer within the bore of an EVG EV501 wafer bonding apparatus. A 2" stainless steel pressure plate used for applying a uniform force to the stack was then placed on top of the stack. Following an evacuation of the bonding chamber to a pressure of approximately $5\times 10^{-4}$ Torr, a force between 3-6 $N/mm^2$ was applied to the stack for 10 min. After this process, the sample stack was removed from the bonder and the tape was peeled from the SiC wafer, thereby removing approximately 90% of the EG layers from the SiC as measured by both Raman and XPS. The tape with the removed EG layers was placed on the prepared $SiO_2$ on Si 'handle' substrate and was then returned to the bonding apparatus, underneath the 2" stainless steel pressure plate, where a force of 3-6 $N/mm^2$ was applied for 10 min. The stack was then removed and placed on a hot plate, where the surface temperature was stabilized at a temperature 1-2$^oC$ above the $120^oC$ release temperature of the tape. This thermal treatment eliminates the adhesion strength of the tape. The tape was removed, leaving behind the transferred EG film on the handle substrate. The tape residue was dissolved using a solution of 1:1:1 toluene:methanol:acetone and a final anneal at $250^oC$ for 10 min was performed to improve the transferred EG to SiO$_2$ adhesion. A schematic of this process is depicted in the Supplemental Information section.

EG film thicknesses were estimated by measuring the attenuation of substrate Raman signal intensity (777 and 964 cm$^{-1}$ for EG on SiC; 521 cm$^{-1}$ for EG transferred to SiO$_2$ on Si) induced by the presence of the EG film.[19] Atomic force microscopy of various regions of interest was used to calibrate the attenuated Raman signal intensity to the measured EG film thickness, fitting the results to the following:

$$I = I_0 e^{-2\alpha d} \quad (1)$$

where $I$ and $I_0$ represent the measured Raman intensity in the presence and absence of an EG film, respectively. The fitting parameter $\alpha$ corresponds to the relative extinction coefficient of the EG films on SiC, ($\alpha$ =0.2 nm$^{-1}$) and of the EG and SiO$_2$ films on Si, ($\alpha$ =0.0576 nm$^{-1}$). Raman measurements were performed confocally using either a 514.5 or a 532 nm laser line focused through a 100X, 0.9 N.A. or a 50X, 0.42 N.A. objective, respectively, in both cases providing a sub-micron laser spot with a power at the sample of approximately 10 mW. The collected signal was passed through an appropriate wavelength Semrock long-pass filter and was detected using either an Ocean Optics QE65000 CCD spectrometer or a half-meter Acton single spectrometer (SpectraPro-2500i) with a Princeton Instruments nitrogen-cooled, back-thinned, deep-depleted CCD array detector (SPEC-10:400BR/LN). Spatial mapping of the Raman peak characteristics (position, full-width at half max, center-of-mass, intensity and calculated EG film thickness) was achieved by translating the sample with respect to the laser spot.

In order to determine the effect of the transfer process on the electrical characteristics of the transferred EG films, Hall effect mobility and carrier density measurements were carried out at 300 K using a van der Pauw configuration. For 10 mm x 10 mm samples, beryllium-copper pressure clips were used as probe contacts to the corners of the as-grown and transferred films. For patterned 200 μm x 200 μm van der Pauw squares, standard probe manipulators were used as current and voltage leads. Measurement currents ranged from 1 to 100 μA while the magnetic field was approximately 2kG.

Presented in Fig. 1 are two-dimensional Raman thickness maps, using the characterization technique described above, for an (a) as-grown EG film on C-face 4H-SiC, (b) an EG film transferred using 3 N/mm$^2$ bonding force onto a SiO$_2$ on Si substrate and (c) a residual EG film on C-face 4H-SiC following the successful removal of the upper EG layers. The map of the as-grown EG film [Fig. 1 (a)], indicates that there is a some considerable variation in the initial film thickness, with the average thickness of this sample found to be approximately 19 nm, with EG film thicknesses typically ranging from 10-30 nm. In comparison, the transferred EG films had average thicknesses ranging from 8-14 nm. Similar variations in the average film thickness of the as-grown and transferred films were observed. Due to these comparable thickness variations observed before and after the EG transfer process, it was difficult to determine if any changes in the thickness uniformity were induced via the transfer. We also determined that the thickness of the residual EG layer on SiC following the transfer ranged from 0-5 nm in thickness, implying that about 87% of the EG film was removed via the transfer procedure. Subsequent XPS measurements verified this transfer efficiency.

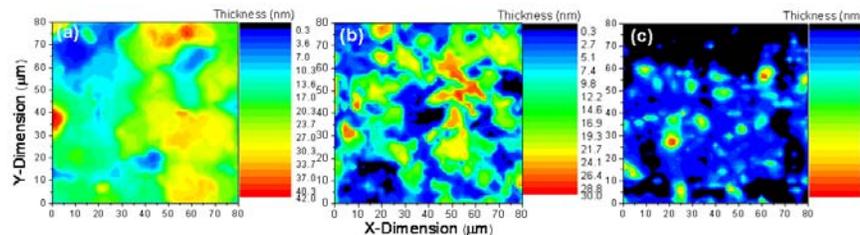

**Figure 1:** Spatial maps of the film thickness for (a) an as-grown EG film on SiC, (b) an EG film transferred using a 3 N/mm$^2$ bonding force onto a SiO$_2$ on Si handle substrate and (c) the residual EG film remaining on SiC after the transfer process. The film thickness was measured using the method outlined in Ref. [19] and calibrated using AFM.

In Fig. 1(b), holes on the order of 5-10 μm in size can be observed in the transferred EG films, which are not present in the as-grown films as illustrated in Fig. 1 (a). In Fig. 1 (c), similar sized regions of thick EG were found remaining on the SiC surface after the transfer process. Presumably, these regions of missing EG within the transferred films correlate to the thick residual EG regions on the SiC after the transfer. In the as-grown EG films, small regions with 10-20 nm reductions in EG film thickness can also be observed. It is possible that the small holes in the transferred films are due to the inability of the bonding force to fully press the transfer tape into these regions where steep reductions in film thickness are found, thus leaving the observed void in the transferred EG films.

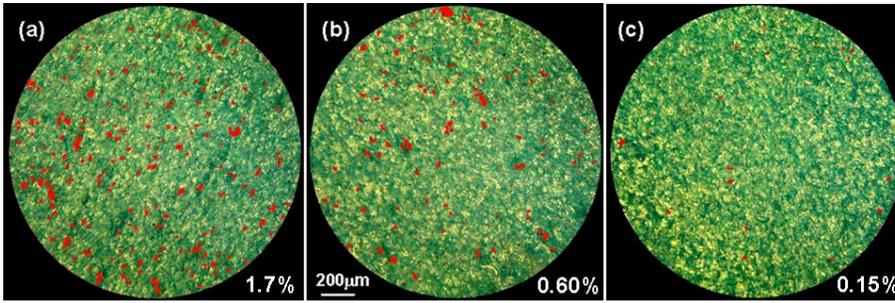

**Figure 2:** Nomarski micrographs of EG films transferred to SiO$_2$ on Si substrates following the application of (a) 3, (b) 4 and (c) 5 N/mm$^2$ force during the bonding of the thermal release tape to the EG on SiC prior to removal. Regions of missing EG have been highlighted in red for clarity and the percentage of the field-of-view where graphene is missing is reported in the lower left hand corner of the images.

In order to optimize the transfer process, a bonding force dependence study was performed. The force applied to the tape/EG/SiC stack was varied from 3-6 N/mm$^2$, with a constant 3 N/mm$^2$ force applied during the second bonding step (tape/EG/handle substrate stack). Nomarski micrographs of the films transferred using the 3, 4 and 5 N/mm$^2$ bonding forces were collected through a 10X, 0.25 NA objective and are presented in Fig. 2 (a)-(c), respectively. These images illustrate that as the force is increased from 3 to 5 N/mm$^2$, the number of regions of missing EG (highlighted in red) in the transferred material are reduced. Therefore increasing the force during the tape to EG bonding stage leads to a more complete and continuous EG film. Subsequent particle analysis studies were performed to measure the percentage of the microscope field-of-view where EG was missing and the corresponding values are presented in Fig. 2 (a)-(c). This was completed using the 'analyze particles' function of ImageJ.[20] These measurements further illustrate the improvement in the transferred EG film quality with increasing bonding force, as the areal percentage of missing EG was reduced ten-fold as the pressure was increased from 3 to 5 N/mm$^2$. Presented in Fig. 3 (a) and (b) are µ-Raman spectra collected from the EG films transferred using 3 and 5 N/mm$^2$ bonding force, respectively. These spectra depict the relative intensities of the 'D' (~1380 cm$^{-1}$), 'G' (~1530 cm$^{-1}$) and '2D' (~2700 cm$^{-1}$) Raman lines from selected areas with the highest (red trace) and lowest (blue trace) D:G intensity ratios $(I_D:I_G)$. Previously, it was reported that this ratio varies inversely with the planar correlation length (approximate grain size) of the graphitic planes and may be used as a qualitative figure of merit for EG films, with a low $I_D:I_G$ indicating a higher quality graphene film.[21] Corresponding spatial maps of the $I_D:I_G$ ratio over a 20 µm x 20 µm area from these EG films. A significant reduction in the $I_D:I_G$ ratio was observed in films transferred using the higher bonding force, thereby indicating a dramatic improvement in the material quality with increasing bonding force. Furthermore, the spatial uniformity of the $I_D:I_G$ ratio was significantly increased in the film transferred using a force of 5 N/mm$^2$. The average values of the $I_D:I_G$ ratio were 0.037 +/- 0.008 and 0.010 +/- 0.002 for the films transferred using 3 and 5 N/mm$^2$, respectively, with the error reported pertaining to the 95% confidence interval. By comparison, $I_D:I_G$ ratios of the as-grown EG films from the same wafer were found to range from 0.009 to to 0.015, indicating that the optimized transfer process induced minimal additional degradation to the EG films. Upon increasing the pressure to 6 N/mm$^2$ (not shown), the uniformity of the transferred film was lost, as small islands of transferred EG were observed instead of a continuous EG film. This force dependence study indicates that a bonding force of approximately 5 N/mm$^2$ is optimal for enabling the transfer of the most continuous, uniform and lowest defectivity EG films from SiC using this thermal release tape. However, efforts involving graphene transfers from different growth substrates and/or different adhesives would require a separate optimization study. In subsequent experiments it was also determined that increasing the bonding force to the tape/EG/handle substrate stack to 5 N/mm$^2$ also improved the efficiency of the transfer, as illustrated by a further improved transfer efficiency of the EG films onto the handle substrates.

To determine the amenability of this transfer process to various other handle substrates, the optimized process described above, with the exception of the handle substrate surface preparation steps, was used to transfer EG films from C-face SiC onto p- and n-type GaN and also onto $Al_2O_3$ films deposited on c-face sapphire via ALD. In the case of the former, the ALD $Al_2O_3$ film was required to modify the highly unreactive surface of the sapphire substrate to enable a successful transfer. The transferred EG films optically appeared very similar to those transferred onto $SiO_2$ using the same bonding force, therefore indicating that the transfer process is relatively insensitive to the handle substrate material.

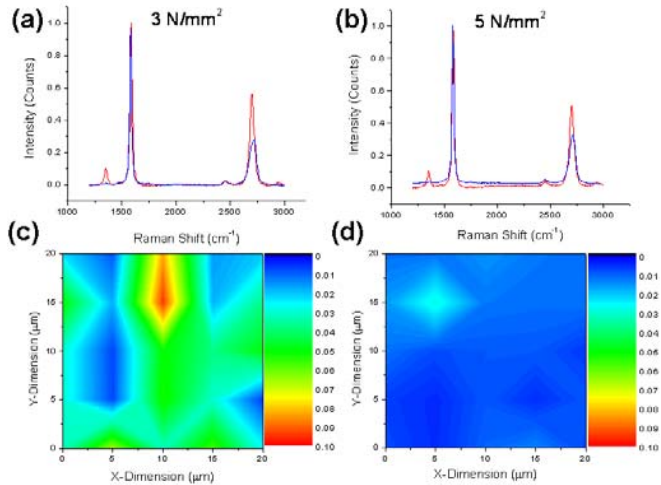

**Figure 3:** µ-Raman spectra collected from EG films transferred to SiO2 on Si substrates using a bonding force of (a) 3 and (b) 5 N/mm². The spectra were collected from regions where there was high (red trace) and low (blue trace) intensities of the Raman 'D' line. Corresponding maps representing the spatial distribution of the ratio of the Raman 'D' to 'G' intensities are presented in (c) and (d), respectively.

In order to determine if any degradation to the electrical properties of the transferred films occurred during the transfer process, Hall measurements were initially performed on as-grown EG films and were repeated following the transfer of these films to $SiO_2$. Room temperature carrier mobilities from the 16 mm x 16 mm as-grown EG material ranged from 909 to 1875 $cm^2$/Vs, with an average mobility of 1485 $cm^2$/Vs being observed, corresponding with an average carrier density of 15.8 x $10^{13}$ $cm^{-3}$. Following the transfer process, the mobility and carrier density from these large-area films both decreased significantly, with post-transfer mobility values ranging from 188 to 269 $cm^2$/Vs, with an average of 201 $cm^2$/Vs, while the carrier density was reduced three-fold to an average of 5.10 x $10^{13}$ $cm^{-2}$. Because this reduction in mobility was observed despite a corresponding decrease in carrier density, it is apparent that this mobility reduction is due to the introduction of additional scattering centers or defects within the EG films during the transfer process. However, even this reduced mobility is still orders of magnitude higher than amorphous silicon or most flexible conductive films, such as organic thin film transistors.[22]

A large reduction in the carrier density in comparison to the values initially measured in the as-grown EG films was observed via the Hall effect following the transfer process. Sun et al.[23] showed that within a C-face EG film that there are layers with varied levels of doping and they further proposed that the highest doped layers are located at or near the SiC-EG interface. As the transfer procedure discussed here leaves behind only the bottom most layers, the reduction in carrier density observed in the transferred EG films is consistent with the upper EG layers having fewer carriers than those layers closest to the SiC interface. One would further expect that the residual EG films on SiC would exhibit a carrier density that was unchanged from the initial as-grown EG film measurements. However, due to a lack of electrical continuity of the post-transfer, residual EG films, Hall measurements were inconclusive. XPS measurements of EG films bound to the thermal-release tape prior to transfer, revealed the presence of a Si 2p peak within the EG films. Subsequent measurements illustrated that the tape was not the source of this Si peak. Therefore, it is possible that these intercalated Si atoms are acting as an isoelectronic dopant and may account, at least in part, for the high sheet carrier densities found in EG in comparison to the reported intrinsic levels.[24, 25]

Previous reports[16] have illustrated that fabrication of smaller area devices (10–200 µm on a side) within EG material has enabled the observation of significantly higher carrier mobilities. In an effort to determine if such increases in carrier mobility would be observed within the transferred EG films and to determine the suitability of these films for electrical device fabrication, one of the films transferred using a force of 3 N/mm² was patterned into a series of device structures including multiple 200 µm x 200 µm van der Pauw squares with gold contact pads. Presented in Fig. 4 (a) is an optical micrograph of one such device. Note the small region in the upper left where EG is not present. Subsequent Raman spatial thickness maps [Fig. 4 (b)] were performed and highlight the locations of the missing EG (shown as black in the spatial map) and provide a measured average EG film thickness of approximately 10 nm.

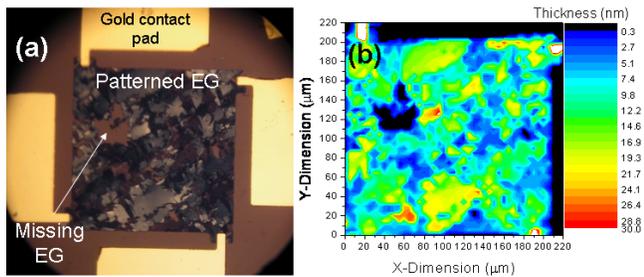

**Figure 4:** (a) Optical image of an EG film patterned into a 200 μm x 200 μm van der Pauw square and (b) the corresponding thickness map of this patterned device determined using the Raman technique outlined in the text.

Despite these van der Pauw squares typically being incomplete films, the mobilities were still found to improve relative to the large-area transferred films, with values ranging from 384 to 803 cm$^2$/Vs, with an average of 739 cm$^2$/Vs, being recorded. These values were therefore approaching 50% of the initial mobilities measured in the as-grown EG films.

Here we have outlined a method enabling the dry transfer of large-area C-face EG films from SiC onto an arbitrary handle substrate, thereby greatly increasing the flexibility of graphene films for most electronic, optoelectronic and mechanical applications. While we have focused this discussion on the transfer of EG onto SiO$_2$ on Si substrates, successful transfers onto both p- and n-type MOCVD GaN and thin ALD-deposited Al$_2$O$_3$ films were also reported. Using this process, patterned devices were fabricated that retained up to 50% of the as-grown carrier mobilities, while a large reduction in the carrier density and the presence of Si within the EG films was observed. These transferred EG films are suitable for optically transparent, conductive films for optoelectronic applications and the observed mobilities are several orders of magnitude higher than either amorphous silicon or most flexible electronic devices, such as organic thin film transistors.[22]

Acknowledgements: We would like to thank Nitto Denko America (Fremont, CA), Electronics Process Material Group (system.nda@nitto.com) for providing the thermal release tape and expertise in the associated transfer and cleaning processes. The authors also express their appreciation to Mr. Steven Binari for the use of his Hall effect measurement system and Dr. Jeremy Robinson for helpful discussions. This work was supported in part by the Office of Naval Research. Support for JLT and JKH was provided by the ASEE.

**Supporting Information**

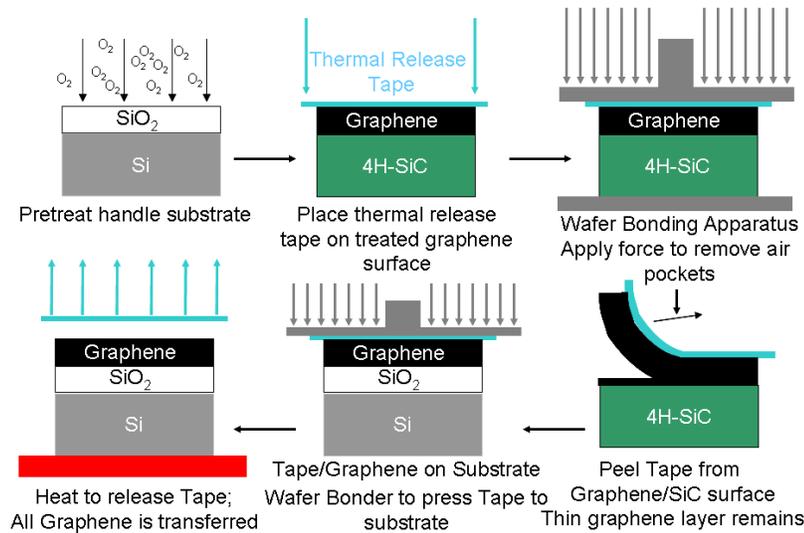

A schematic detailing the transfer procedure is presented in Supplemental Fig. 1.

1. Pretreat the handle substrate using a standard chemical clean (if necessary) and an oxygen plasma for improved surface bonding to graphene transfer.

2. The thermal release tape is placed on the graphene surface, forming the graphene/tape stack.

3. The graphene/tape stack is placed within the bonding apparatus and a defined force is applied to the stack.

4. The stack is removed from the bonding apparatus and the tape, along with the majority of the EG layers, are peeled from the SiC substrate.

5. The EG bonded to the tape is placed on the 'handle' substrate (e.g. $SiO_2$ on Si) and is placed in the bonding apparatus where a similar force is applied to the EG on 'handle' substrate stack.

6. The EG on 'handle' substrate stack is removed and is heated on a hot plate maintained at a temperature greater than the release temperature of the thermal release tape used. The tape, after losing its adhesive properties is then removed from the sample surface. A $250^{o}C$ hotplate anneal in atmosphere and solvent clean may be used to improve graphene/substrate adhesion and remove the tape residue, respectively.

SYNOPSIS TOC